\documentclass{article}
\usepackage{dcase2024,amsmath,graphicx,url,times,booktabs, tabularx}

\title{Self Training and Ensembling Frequency Dependent Networks \\ with Coarse Prediction Pooling and Sound Event Bounding Boxes}

\name{Hyeonuk Nam,
      Deokki Min,
      Seungdeok Choi, 
      Inhan Choi,
      Yong-Hwa Park$\sthanks{This work was supported by the Institute of Civil Military Technology Cooperation funded by the Defense Acquisition Program Administration and Ministry of Trade, Industry and Energy of Korean government under grant No. UM22409RD4, and Korea Research Institute of Ships and Ocean engineering a grant from Endowment Project of “Development of Open Platform Technologies for Smart Maritime Safety and Industries” funded by Ministry of Oceans and Fisheries(PES5230).}$
      }
\address {Korea Advanced Institute of Science and Technology, South Korea,\\
        \{frednam, minducky, haroldchoi6, ds5amk, yhpark\}@kaist.ac.kr
 }

\begin{document}
\ninept
\maketitle

\begin{sloppy}

\begin{abstract}
To tackle sound event detection (SED), we propose \textit{frequency dependent networks (FreDNets)}, which heavily leverage frequency-dependent methods. We apply frequency warping and FilterAugment, which are frequency-dependent data augmentation methods. The model architecture consists of 3 branches: audio teacher-student transformer (ATST) branch, BEATs branch and CNN branch including either partial dilated frequency dynamic convolution (PDFD conv) or squeeze-and-Excitation (SE) with time-frame frequency-wise SE (tfwSE). To train MAESTRO labels with coarse temporal resolution, we applied max pooling on prediction for the MAESTRO dataset. Using best ensemble model, we applied self training to obtain pseudo label from DESED weak set, unlabeled set and AudioSet. AudioSet pseudo labels, filtered to focus on high-confidence labels, are used to train on DESED dataset only. We used change-detection-based sound event bounding boxes (cSEBBs) as post processing for ensemble models on self training and submission models. The resulting FreDNet was ranked 2nd in DCASE 2024 Challenge Task 4.
\end{abstract}

\begin{keywords}
frequency dynamic convolution, audio pre-trained models, coarse prediction pooling, label filtering, sound event bounding boxes
\end{keywords}

\section{Introduction}
In this work, we address the problem of sound event detection (SED) with heterogeneous datasets, including Domestic Environment Sound Event Detection (DESED) and Multi-Annotator Estimated STROng labels (MAESTRO) \cite{dcase2024baseline, DCASEtask4, maestro}. Since SED is a very delicate task requiring classification with time localization, the difference between two datasets must be carefully addressed. While DESED uses hard labels with fine temporal resolution (base unit of one millisecond) and includes ten target sound events those occur in domestic environment, MAESTRO uses soft labels representing confidence with coarse temporal resolution (base unit of one second) and includes seventeen target sound events those occur on outdoor environments. There are only few target sound events overlapping. For the target sound events those do not overlap, target sound events from one dataset might exist in the other dataset but they are not explicitly labeled. This arouses the problem of potentially missing labels \cite{dcase2024baseline}. To tackle this problem, DCASE2024 Challenge Task 4 baseline is designed to train both datasets using single model architecture to output for 27 classes, while masking the classes from one dataset when training for the other dataset \cite{dcase2024baseline}.

Our primary approach is to build strong classifier that works on both datasets. To achieve this, we applied two frequency-dependent data augmentations: frequency warping and FilterAugment \cite{atstsed, filtaug}. Then, we applied advanced variants of frequency dynamic convolution (FDY conv) to CNN branch of the baseline \cite{FDY, DFD, PFD}. We also used squeeze and excitation (SE) with time-frame frequency wise SE (tfwSE) to CNN branch \cite{freqatt}. In addition to CNN and BEATs branch, we added audio teacher student transformer (ATST) branch to form three-branched models \cite{atstsed, beats, ATST}. In order to match the granularity of strong prediction tailored for DESED to MAESTRO strong labels, we pooled strong predictions. Since frequency-dependent methods are heavily used, we call above network architecture as \textit{Frequency Dependent Networks (FreDNets)}. We used change-detection-based sound event bounding boxes (cSEBBs) as post processing \cite{SEBB}. With ensemble of FreDNets post-processed by cSEBBs, we produced pseudo labels on AudioSets, and used them to train new FreDNets \cite{audioset}.

The main contributions of this paper are as follows:
\vspace{-5pt}
\begin{enumerate}
    \itemsep-0.5em
    \item{Proposed \textit{Frequency dependent networks (FreDNets)} heavily utilizes frequency-dependent methods to outperform the baseline by 15.1\% without ensemble.}
    \item{Proposed coarse prediction pooling successfully harmonizes the temporal resolution difference between the datasets.}
    \item{Partial dilated frequency dynamic convolution (PDFD conv) is lighter than FDY conv or DFD conv providing various models, thus proven to be advantageous upon ensemble.}
\end{enumerate}

\section{Methods}
\subsection{Frequency-Dependent Data Augmentations}
\vspace{-8pt}
In addition to mixup applied in the baseline \cite{dcase2024baseline, mixup}, we added frequency warping and FilterAugment \cite{atstsed, filtaug}. The sequence of operation is as follows: mixup, frequency warping then FilterAugment. Frequency warping is random resize crop applied only along frequency dimension to zoom into frequency dimension with random proportion. As it also works as frequency shift, we did not apply frequency shift. Then, we applied linear type FilterAugment with dB range from -3 dB to +3 dB. This is narrower range compared to the setting in \cite{filtaug}. FilterAugment applies random weights over different frequency ranges to simulate different acoustic environments. Data augmentation is only applied to CNN branch as shown at the top of Fig. \ref{fig:framework}, because the other two branches are not trainable. %Since frequency dependency is an important issue in SED, these two methods showed performance gain upon simultaneous application.

\begin{figure*}[ht]
\centerline{\includegraphics[width=17cm]{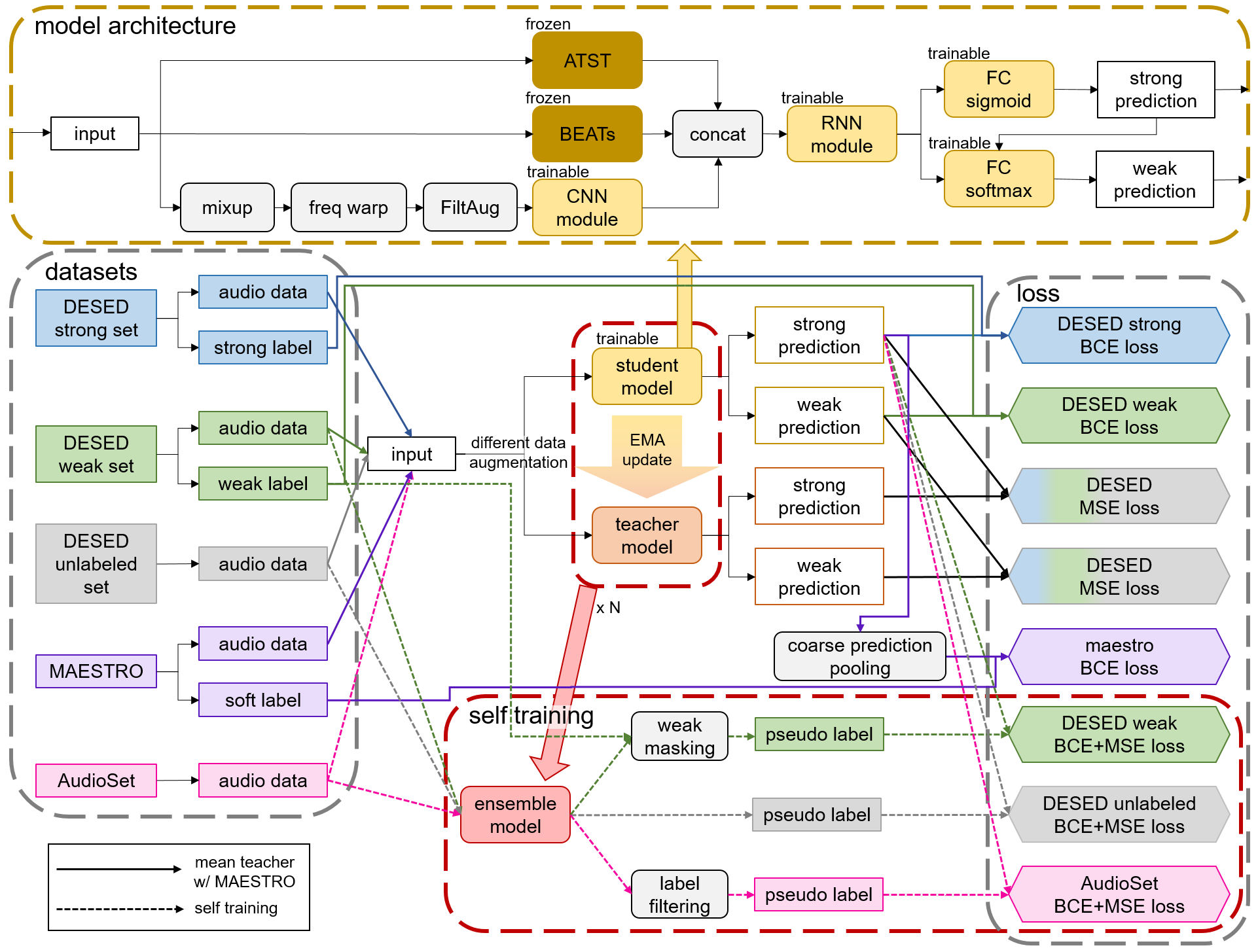}}
\vspace{-10pt}
\caption{An illustration of framework for training and self training FreDNets.}
\label{fig:framework}
\vspace{-10pt}
\end{figure*}

\begin{figure}[t]
\centerline{\includegraphics[width=8.5cm]{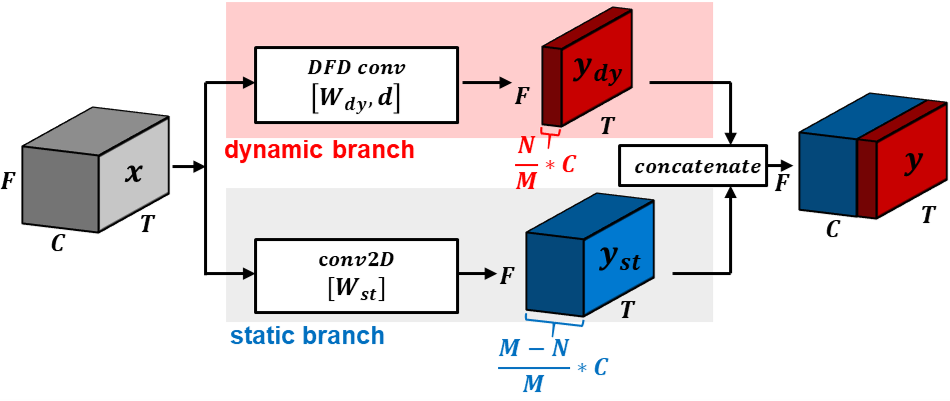}}
\vspace{-10pt}
\caption{An illustration of partial dilated frequency dynamic convolution (PDFD conv). It involves a dynamic DFD conv branches and a static 2D convolution branch.}
\label{fig:MDFDconv}
\vspace{-12pt}
\end{figure}

\subsection{Frequency-Dependent CNN Methods}
\vspace{-8pt}
To further enhance the capacity of the network, CNN and RNN channels are doubled. Either variants of frequency dynamic convolution (FDY conv) or squeeze-and-excitation (SE) are used to make CNN modules leverage frequency-dependent attention methods.

FDY conv applies frequency-adaptive convolution kernel to release translational equivariance along frequency axis of time-frequency features \cite{FDY}. To lighten FDY conv, we applied partial frequency dynamic convolution (PFD conv) with proportion of one over eight \cite{PFD}. To expand and diversify the basis kernels, we applied dilated frequency dynamic convolution (DFD conv), which applies frequency-wise dilation to four basis kernels of PFD conv. We refer to this method as partial dilated frequency dynamic convolution (PDFD conv), which is illustrated in Fig. \ref{fig:MDFDconv}. Using different dilation sizes to PDFD conv resulted in various models which are advantageous on model ensemble \cite{dcase2022mytechrep}. While multi-dilated frequency dynamic convolution (MDFD conv) yields in the best performance, we used PDFD since it offers best cost-performance balance \cite{PFD}. In addition to PDFD conv, we also used SE with time-frame frequency-wise SE (tfwSE) for model variety upon ensemble \cite{freqatt, dcase2022mytechrep}.

\subsection{Transformer-based Pre-trained Audio Models}
\vspace{-8pt}
In addition to CNN branch, two transformer-based pre-trained audio models are used: BEATs and ATST-frame. Frame-wise feature of BEATs and ATST-frame are used to optimally enhance SED which needs to give frame-wise predictions. Embeddings extracted for both methods are pooled into same frame size as output by CNN module output, then concatenated with the output from CNN module along channel dimension, and then processed by fully connected layers along channel dimension. Then the output is fed to RNN module. Note that since transformer-based Audio models divide mel spectrogram into patches and then apply positional encoding to the patches, they implicitly apply frequency-dependent processing. Thus two audio models can be regarded frequency-dependent methods as well. Fine tuning of ATST is not used in this work as it negatively affects MPAUC on MAESTRO \cite{atstsed}.

\subsection{Coarse Prediction Pooling}
\vspace{-8pt}
In order to address the different temporal resolution of DESED and MAESTRO, we applied coarse prediction pooling for MAESTRO data. While FreDNets' predictions have temporal resolution of 64ms per frame (156 frames for 10 seconds), MAESTRO label has temporal resolution of 1s per frame (10 frames for 10 seconds). To make fine predictions into coarse predictions, we apply max pooling on FreDNets' MAESTRO prediction. To be more specific, we zero-padded 2 frames before and after the prediction and max pooled with filter size and stride of 16. Although this is not precise pooling, this choice was made to quickly and simply implement the idea.

\subsection{Sound Event Bounding Boxes}
\vspace{-8pt}
Polyphonic sound detection score (PSDS) applies various thresholds to the SED prediction to obtain threshold-independent evaluation values \cite{PSDS, truePSDS}. However, as threshold differs, onset and offset of sound events also varies. To make onset and offset of sound events independent of the thresholds, sound event bounding boxes (SEBBs) are proposed to combine confidence values with very fine onset and offset values into representative confidence, onset and offset values \cite{SEBB}. In this work, change-detection-based SEBBs (cSEBBs) are used.

\subsection{Self Training using AudioSet}
\vspace{-8pt}
To obtain pseudo labels on DESED weak set, DESED unlabeled set and AudioSet, we used ensemble of FreDNet using PDFD-CNN modules with varying dilation size sets, SE+tfwSE-CNN and PFD-CNN with varying seeds and then applied cSEBBs \cite{dcase2022mytechrep, dcase2023a_1st}. As DESED weak set is given with weak labels, pseudo label for weak set is masked with given weak labels as in \cite{mytechreport}. Since AudioSet has inconsistent label quality, we applied self training on whole dataset to obtain confidence from our ensemble FreDNet. For AudioSet, we filtered data files having pseudo label values (confidence) above 0.7 on 27 target events to focus on labels with high confidence. we discarded event labels with confidence value below 0.01 to reduce pseudo label metadata size, and removed the files of which events above 0.7 are only composed of subset of (speech, people talking, children voices) to reduce the data imbalance toward speech. The count of filtered AudioSet files is 153,977.

Upon use of AudioSet pseudo label, both soft label and hard label obtained by thresholding with 0.5 are used to train SED model. For mean square error (MSE) loss and binary cross entropy (BCE) loss are used for soft and hard labels respectively as shown in red dashed line box in Fig. \ref{fig:framework}. Only 10 target sound events for DESED are trained using filtered AudioSet as it degraded MPAUC when trained on MAESTRO target sound events, although it was meant to train on 17 target sound events in MAESTRO as well.

\subsection{Ensemble}
\vspace{-8pt}
Ensemble model averaged predictions from various models. To maximize the effect of ensemble, we used different models including PFD-CRNN, PDFD-CRNN with different dilation size sets, SE+tfwSE-CRNN, and PFD-CRNN with different seeds. For each model setting, the student and teacher models with the best sum score (PSDS1+MPAUC) are selected for ensemble. The model combinations used for each ensemble setting is shown in Table \ref{tab:ensemble}. Ensemble 1 is used to extract pseudo labels from AudioSet. Ensemble 2 and 3 are used for DCASE Challenge submission. While PFD-CRNNs with different seeds are generally worse than models with seed of 42, models with different seeds do help enhancing ensemble performance.

\begin{table}[t]
\caption{Components models of ensemble models. $1/8$ denotes that $1/8$ of PFD conv or PDFD conv output channel is from FDY conv or DFD conv. Sd, ds and st implies seed, dilation sizes and self training. For model names, CRNN is omitted for brevity.}
\centering
\begin{tabular}{c|c}
\hline
\textbf{ensemble} & \textbf{models}  \\ 
\hline
1               & PFD(1/8), PFD(1/8, sd=2),\\
                & PFD(1/8, sd=12), PFD(1/8, sd=16), \\
                & PFD(1/8, sd=27), PFD(1/8, sd=34), \\
                & PDFD(1/8, ds=1/1/2/2), PDFD(1/8, ds=1/1/3/3), \\
                & PDFD(1/8, ds=2/2/3/3), PDFD(1/8, ds=1/1/2/3), \\
                & PDFD(1/8, ds=1/2/2/3), PDFD(1/8, ds=1/2/3/3) \\
\hline
2               & PFD(1/8), PFD(1/8, sd=16), \\
                & PDFD(1/8, ds=1/1/2/2), PDFD(1/8, ds=1/1/3/3), \\
                & PDFD(1/8, ds=1/1/2/3), PDFD(1/8, ds=1/2/2/3), \\
                & PDFD(1/8, ds=1/2/3/3), \\
                & st-PFD(1/8), st-PFD(1/8, sd=2), \\
                & st-PFD(1/8, sd=12), st-SE+tfwSE,\\
                & st-PDFD(1/8, ds=1/1/2/2), st-PDFD(1/8, ds=1/1/2/3), \\
                & st-PDFD(1/8, ds=1/2/2/3), st-PDFD(1/8, ds=1/2/3/3) \\
\hline
3               & PFD(1/8), PFD(1/8, sd=16), SE+tfwSE, \\
                & PDFD(1/8, ds=1/1/2/2), PDFD(1/8, ds=1/1/3/3), \\
                & PDFD(1/8, ds=1/1/2/3), PDFD(1/8, ds=1/2/2/3), \\
                & PDFD(1/8, ds=1/2/3/3), \\
                & st-PFD(1/8), st-PFD(1/8, sd=2), \\
                & st-PFD(1/8, sd=12), st-PFD(1/8, sd=27), \\
                & st-SE+tfwSE,\\
                & st-PDFD(1/8, ds=1/1/2/2), st-PDFD(1/8, ds=1/1/2/3), \\
                & st-PDFD(1/8, ds=1/2/2/3), st-PDFD(1/8, ds=1/2/3/3), \\
\hline
\end{tabular}
\label{tab:ensemble}
\vspace{-12pt}
\end{table}

\section{Experimental Settings}
 \begin{table*}[ht]
\caption{Performance of FreDNets.}
\centering
\begin{tabular}{c|ccc|ccc|c}
\hline
\textbf{models}      & \textbf{pre-trained models} & \textbf{post-processing} & \textbf{self training} & \textbf{PSDS1} & \textbf{MPAUC} & \textbf{sum}   & \textbf{\# submission} \\
\hline
Baseline \cite{dcase2024baseline} & BEATs          & median filter            & -                      & 0.520          & 0.637          & 1.157          & -                      \\
PFD-CRNN(1/8)        & ATST  + BEATs               & median filter            & -                      & 0.516          & 0.775          & 1.293          & -                      \\
PFD-CRNN(1/8, sd=2)  & ATST  + BEATs               & median filter            & -                      & 0.502          & 0.766          & 1.268          & -                      \\
PFD-CRNN(1/8, sd=12) & ATST  + BEATs               & median filter            & -                      & 0.514          & 0.765          & 1.279          & -                      \\
PFD-CRNN(1/8, sd=16) & ATST  + BEATs               & median filter            & -                      & 0.514          & 0.772          & 1.286          & -                      \\
PFD-CRNN(1/8, sd=27) & ATST  + BEATs               & median filter            & -                      & 0.514          & 0.763          & 1.277          & -                      \\
PFD-CRNN(1/8, sd=34) & ATST  + BEATs               & median filter            & -                      & 0.508          & 0.769          & 1.276          & -                      \\
PDFD-CRNN(1/8, 1122) & ATST  + BEATs               & median filter            & -                      & 0.519          & 0.773          & 1.292          & -                      \\
PDFD-CRNN(1/8, 1133) & ATST  + BEATs               & median filter            & -                      & 0.523          & 0.767          & 1.290          & -                      \\
PDFD-CRNN(1/8, 2233) & ATST  + BEATs               & median filter            & -                      & 0.515          & 0.772          & 1.287          & -                      \\
PDFD-CRNN(1/8, 1123) & ATST  + BEATs               & median filter            & -                      & 0.518          & \textbf{0.776} & 1.294          & -                      \\
PDFD-CRNN(1/8, 1223) & ATST  + BEATs               & median filter            & -                      & \textbf{0.526} & 0.772          & \textbf{1.298} & -                      \\
PDFD-CRNN(1/8, 1233) & ATST  + BEATs               & median filter            & -                      & 0.518          & 0.774          & 1.292          & -                      \\
SE+tfwSE-CRNN        & ATST  + BEATs               & median filter            & -                      & 0.507          & 0.773          & 1.280          & -                      \\
\hline
Ensemble 1           & ATST  + BEATs               & median filter            & -                      & \textbf{0.527} & \textbf{0.790} & \textbf{1.317} & -                      \\
\hline
Ensemble 1           & ATST  + BEATs               & cSEBBs                   & -                      & \textbf{0.577} & \textbf{0.790} & \textbf{1.367} & -                      \\
\hline
PFD-CRNN(1/8)        & ATST  + BEATs               & median filter            & True                   & \textbf{0.539} & 0.773          & \textbf{1.312} & -                      \\
PFD-CRNN(1/8, sd=2)  & ATST  + BEATs               & median filter            & True                   & 0.534          & 0.766          & 1.300          & -                      \\
PFD-CRNN(1/8, sd=12) & ATST  + BEATs               & median filter            & True                   & 0.534          & 0.753          & 1.287          & -                      \\
PFD-CRNN(1/8, sd=27) & ATST  + BEATs               & median filter            & True                   & 0.531          & 0.750          & 1.287          & -                      \\
PDFD-CRNN(1/8, 1122) & ATST  + BEATs               & median filter            & True                   & 0.530          & 0.774          & 1.304          & -                      \\
PDFD-CRNN(1/8, 1133) & ATST  + BEATs               & median filter            & True                   & 0.535          & 0.761          & 1.296          & -                      \\
PDFD-CRNN(1/8, 1123) & ATST  + BEATs               & median filter            & True                   & 0.537          & \textbf{0.775} & \textbf{1.312} & -                      \\
PDFD-CRNN(1/8, 1223) & ATST  + BEATs               & median filter            & True                   & 0.533          & 0.772          & 1.305          & -                      \\
PDFD-CRNN(1/8, 1233) & ATST  + BEATs               & median filter            & True                   & 0.532          & 0.772          & 1.304          & -                      \\
SE+tfwSE-CRNN        & ATST  + BEATs               & median filter            & True                   & 0.525          & 0.767          & 1.292          & -                      \\
\hline
PFD-CRNN(1/8)        & ATST  + BEATs               & cSEBBs                   & True                   & 0.551          & 0.773          & 1.324          & 1                      \\
PDFD-CRNN(1/8, 1123) & ATST  + BEATs               & cSEBBs                   & True                   & \textbf{0.557} & \textbf{0.775} & \textbf{1.332} & 2                      \\
\hline
Ensemble 2           & ATST  + BEATs               & median filter            & True                   & \textbf{0.537} & 0.788          & \textbf{1.325} & -                      \\
Ensemble 3           & ATST  + BEATs               & median filter            & True                   & 0.536          & \textbf{0.789} & \textbf{1.325} & -                      \\
\hline
Ensemble 2           & ATST  + BEATs               & cSEBBs                   & True                   & \textbf{0.575} & 0.788          & \textbf{1.363} & 3                      \\
Ensemble 3           & ATST  + BEATs               & cSEBBs                   & True                   & 0.574          & \textbf{0.789} & \textbf{1.363} & 4                   \\
\hline
\end{tabular}
\label{tab:results}
\vspace{-12pt}
\end{table*}
\subsection{Implementation Details}
\vspace{-8pt}
DESED and MAESTRO data are processed to be 10 seconds clip with 16kHz sampling rate \cite{dcase2024baseline, maestro, dcasebaseline}. Mel spectrogram is used for input feature. The network is composed of three-branched ATST-BEATs-CNN modules which are then fed to RNN module and Fully Connected layers as shown in Fig. \ref{fig:framework}. The Mean Teacher method is employed to train FreDNets using the DESED unlabeled set \cite{dcasebaseline, meanteacher}. Binary cross entropy (BCE) loss is used to train strong prediction for DESED strong set and its strong label, weak prediction for DESED weak set and its weak label, and strong prediction of MAESTRO and its soft label. Note that strong prediction goes through coarse label prediction before the loss function to match the granularity of prediction and label. For consistency loss for strong and weak predictions of DESED sets, mean square error (MSE) loss is used. For pseudo labels for DESED weakly labeled set, unlabeled set and AudioSet, both BCE and MSE losses are used. Default seed is set to 42. GPU used for training is NVIDIA RTX A6000. For post-processing, we use either cSEBBs or a median filter as reported in Table \ref{tab:results}. The median filter refers to class-independent 7-frames-sized median filter.

\subsection{Evaluation Metrics}
\vspace{-8pt}
True PSDS1 was used to evaluate SED performance on DESED \cite{PSDS, truePSDS}. While previous DCASE challenge task 4 used two types of PSDS (PSDS1 favoring time localization and PSDS2 favoring accurate classification), only PSDS1 is used in this year as PSDS2 is rather an audio tagging metric \cite{SEBB, mytechreport}. For MAESTRO performance evaluation, MPAUC is used \cite{dcase2024baseline}. We optimized the model based on average score of PSDS1 + MPAUC on 4 independent training runs. The scores reported in the table are from the models with best sum scores among 4 independent training runs within each model setting.

\section{Results}
The results are summarized in Table \ref{tab:results}, highlighting the performance improvements achieved by our proposed methods. The PSDS and MPAUC values are obtained on real validation sets of DESED and MAESTRO respectively. As shown in the results, PFD-CRNN and PDFD-CRNNs do not significantly vary in their performance. However, as their roles differ from each other, ensembling differently dilated PDFD-CRNNs results in decent performance. Likewise, although slightly worse than PDFD-CRNNs, SE-tfwSE-CRNN and PFD-CRNNs with different seeds do help for ensemble. From the results, it could be inferred that use of FreDNet including frequency-wise data augmentation, PDFD conv, BEATs, ATST-frame, coarse prediction pooling enhances MPAUC by large margine while PSDS is not significantly improved. Rather, use of cSEBBs and self training improves PSDS significantly. Final best score without ensemble model outperforms the baseline by 15.1\% and best score with ensemble outperforms the baseline by 18.2\%. While ensemble 1 model slightly outperforms ensemble 2 and 3 those outperformed the baseline by 17.8\%, submission was made with latter two as they contain self-trained models thus are expected to retain better generalization capability.

\section{Conclusion}
In this study, we presented Frequency Dependent Networks (FreDNet) for SED. FreDNet leverages frequency-dependent data augmentation techniques, frequency warping and FilterAugment, and incorporates advanced neural network architectures such as frequency dependent CNNs and transformer-based pre-trained models. Experiments show that the proposed FreDNet architecture, when combined with PDFD conv, SE, and coarse prediction pooling, significantly improves SED performance especially on MPAUC. The use of cSEBBs further enhances performance by refining onset and offset predictions on PSDS. The ensemble models, integrating various FreDNet settings, achieved substantial performance gains over the baseline, with the best ensemble model outperforming the baseline by 18.2\%. Our approach shows promise for robust SED in diverse environments, highlighting the effectiveness of frequency-dependent methods and the importance of ensemble strategies in improving model performance. The model described in this work was ranked 2nd in DCASE 2024 Challenge Task 4.

% -------------------------------------------------------------------------
% Either list references using the bibliography style file IEEEtran.bst
\bibliographystyle{IEEEtran}
\bibliography{refs}

\begin{thebibliography}{10}
\providecommand{\url}[1]{#1}
\def\UrlFont{\rmfamily}
\providecommand{\newblock}{\relax}
\providecommand{\bibinfo}[2]{#2}
\providecommand\BIBentrySTDinterwordspacing{\spaceskip=0pt\relax}
\providecommand\BIBentryALTinterwordstretchfactor{4}
\providecommand\BIBentryALTinterwordspacing{\spaceskip=\fontdimen2\font plus
\BIBentryALTinterwordstretchfactor\fontdimen3\font minus \fontdimen4\font\relax}
\providecommand\BIBforeignlanguage[2]{{%
\expandafter\ifx\csname l@#1\endcsname\relax
\typeout{** WARNING: IEEEtran.bst: No hyphenation pattern has been}%
\typeout{** loaded for the language `#1'. Using the pattern for}%
\typeout{** the default language instead.}%
\else
\language=\csname l@#1\endcsname
\fi
#2}}

\bibitem{dcase2024baseline}
S.~Cornell, J.~Ebbers, C.~Douwes, I.~Martín-Morató, M.~Harju, A.~Mesaros, and R.~Serizel, ``Dcase 2024 task 4: Sound event detection with heterogeneous data and missing labels,'' \emph{arXiv preprint arXiv:2406.08056}, 2024.

\bibitem{DCASEtask4}
N.~Turpault, R.~Serizel, A.~Parag~Shah, and J.~Salamon, ``{Sound event detection in domestic environments with weakly labeled data and soundscape synthesis},'' in \emph{{Workshop on Detection and Classification of Acoustic Scenes and Events}}, 2019.

\bibitem{maestro}
I.~Martín-Morató and A.~Mesaros, ``Strong labeling of sound events using crowdsourced weak labels and annotator competence estimation,'' \emph{IEEE/ACM Transactions on Audio, Speech, and Language Processing}, 2023.

\bibitem{atstsed}
N.~Shao, X.~Li, and X.~Li, ``Fine-tune the pretrained atst model for sound event detection,'' in \emph{International Conference on Acoustics, Speech and Signal Processing (ICASSP)}, 2024.

\bibitem{filtaug}
H.~Nam, S.-H. Kim, and Y.-H. Park, ``Filteraugment: An acoustic environmental data augmentation method,'' in \emph{International Conference on Acoustics, Speech and Signal Processing (ICASSP)}, 2022.

\bibitem{FDY}
H.~Nam, S.-H. Kim, B.-Y. Ko, and Y.-H. Park, ``Frequency dynamic convolution: Frequency-adaptive pattern recognition for sound event detection,'' in \emph{Proc. Interspeech}, 2022.

\bibitem{DFD}
H.~Nam, S.-H. Kim, D.~Min, J.~Lee, and Y.-H. Park, ``Diversifying and expanding frequency-adaptive convolution kernels for sound event detection,'' \emph{arXiv preprint arXiv:2406.05341}, 2024.

\bibitem{PFD}
H.~Nam and Y.-H. Park, ``Pushing the limit of sound event detection with multi-dilated frequency dynamic convolution,'' \emph{arXiv preprint arXiv:2406.13312}, 2024.

\bibitem{freqatt}
H.~Nam, S.-H. Kim, D.~Min, and Y.-H. Park, ``Frequency \& channel attention for computationally efficient sound event detection,'' in \emph{{Workshop on Detection and Classification of Acoustic Scenes and Events}}, 2023.

\bibitem{beats}
S.~Chen, Y.~Wu, C.~Wang, S.~Liu, D.~Tompkins, Z.~Chen, W.~Che, X.~Yu, and F.~Wei, ``Beats: Audio pre-training with acoustic tokenizers,'' in \emph{International Conference on Machine Learning}, 2023.

\bibitem{ATST}
X.~LI and X.~Li, ``Atst: Audio representation learning with teacher-student transformer,'' in \emph{Proc. Interspeech}, 2022.

\bibitem{SEBB}
J.~Ebbers, F.~G. Germain, G.~Wichern, and J.~L. Roux, ``Sound event bounding boxes,'' \emph{arXiv preprint arXiv:2406.04212}, 2024.

\bibitem{audioset}
J.~F. Gemmeke, D.~P.~W. Ellis, D.~Freedman, A.~Jansen, W.~Lawrence, R.~C. Moore, M.~Plakal, and M.~Ritter, ``Audio set: An ontology and human-labeled dataset for audio events,'' in \emph{2017 IEEE International Conference on Acoustics, Speech and Signal Processing (ICASSP)}, 2017.

\bibitem{mixup}
H.~Zhang, M.~Cisse, Y.~N. Dauphin, and D.~Lopez-Paz, ``mixup: Beyond empirical risk minimization,'' in \emph{International Conference on Learning Representations}, 2018.

\bibitem{dcase2022mytechrep}
H.~Nam, S.-H. Kim, D.~Min, B.-Y. Ko, S.-D. Choi, and Y.-H. Park, ``Frequency dependent sound event detection for dcase 2022 challenge task 4,'' DCASE2022 Challenge, Tech. Rep., 2022.

\bibitem{PSDS}
{\c{C}}.~Bilen, G.~Ferroni, F.~Tuveri, J.~Azcarreta, and S.~Krstulovi\'{c}, ``A framework for the robust evaluation of sound event detection,'' in \emph{International Conference on Acoustics, Speech and Signal Processing (ICASSP)}, 2020, pp. 61--65.

\bibitem{truePSDS}
J.~Ebbers, R.~Haeb-Umbach, and R.~Serizel, ``Threshold independent evaluation of sound event detection scores,'' in \emph{International Conference on Acoustics, Speech and Signal Processing (ICASSP)}, 2022.

\bibitem{dcase2023a_1st}
J.~W. Kim, S.~W. Son, Y.~Song, H.~K. Kim, I.~H. Song, and J.~E. Lim, ``Semi-supervised learning-based sound event detection using frequency dynamic convolution with large kernel attention for {DCASE} challenge 2023 task 4,'' DCASE2023 Challenge, Tech. Rep., 2023.

\bibitem{mytechreport}
H.~Nam, B.-Y. Ko, G.-T. Lee, S.-H. Kim, W.-H. Jung, S.-M. Choi, and Y.-H. Park, ``Heavily augmented sound event detection utilizing weak predictions,'' DCASE2021 Challenge, Tech. Rep., 2021.

\bibitem{dcasebaseline}
\BIBentryALTinterwordspacing
N.~Turpault. Dcase2021 task4 baseline. GitHub. Available: https://github.com/DCASE-REPO/DESED\_task. [Online]. Available: \url{https://github.com/DCASE-REPO/DESED\_task}
\BIBentrySTDinterwordspacing

\bibitem{meanteacher}
A.~Tarvainen and H.~Valpola, ``Mean teachers are better role models: Weight-averaged consistency targets improve semi-supervised deep learning results,'' in \emph{Advances in Neural Information Processing Systems}, vol.~30, 2017.

\end{thebibliography}

%
% or list them by yourself
% \begin{thebibliography}{9}
% 
% \bibitem{dcase2016web}
%   \url{http://www.cs.tut.fi/sgn/arg/dcase2016/}.
%
% \bibitem{IEEEPDFSpec}
%   {PDF} specification for {IEEE} {X}plore$^{\textregistered}$,
%   \url{http://www.ieee.org/portal/cms_docs/pubs/confstandards/pdfs/IEEE-PDF-SpecV401.pdf}.
%
% \bibitem{PDFOpenSourceTools}
%   Creating high resolution {PDF} files for book production with 
%   open source tools, 
%   \url{http://www.grassbook.org/neteler/highres_pdf.html}.
%
% \bibitem{eWilliams1999}
% E. Williams, \emph{Fourier Acoustics: Sound Radiation and Nearfield Acoustic
%   Holography}. London, UK: Academic Press, 1999.
% 
% \bibitem{ieeecopyright}
%   \url{http://www.ieee.org/web/publications/rights/copyrightmain.html}.
%
% \bibitem{cJones2003}
% C. Jones, A. Smith, and E. Roberts, ``A sample paper in conference
%   proceedings,'' in \emph{Proc. IEEE ICASSP}, vol. II, 2003, pp. 803--806.
% 
% \bibitem{aSmith2000}
% A. Smith, C. Jones, and E. Roberts, ``A sample paper in journals,'' 
%   \emph{IEEE Trans. Signal Process.}, vol. 62, pp. 291--294, Jan. 2000.
% 
% \end{thebibliography}

\end{sloppy}
\end{document}